# Local tuning of the order parameter in superconducting weak links: A zero-inductance nano device

Roni Winik[1,2], Itamar Holzman[1,3], Emanuele G. Dalla Torre[4], Eyal Buks[2] and Yachin Ivry[1,3,*]

**Affiliations**
1. Solid State Institute, Technion – Israel Institute of Technology, Haifa, 32000, Israel.
2. Andrew and Erna Viterbi Department of Electrical Engineering, Technion, Haifa 32000 Israel.
3. Department of Materials Science & Engineering, Technion – Israel Institute of Technology, Haifa, 32000, Israel.
4. Department of Physics, Bar-Ilan University, Ramat Gan 5290002, Israel

* Correspondence to: ivry@technion.ac.il

**Abstract**

Controlling both the amplitude and phase of the quantum order parameter ($\psi$) in nanostructures is important for next-generation information and communication technologies. The long-range coherence of attractive electrons in superconductors render these materials as a nearly ideal platform for such applications. To-date, control over $\psi$ has remained limited to the macroscopic scale, either by adjusting untunable materials properties, such as film thickness, stoichiometry and homogeneity or by tuning external magnetic fields. Yet, although local tuning of $\psi$ is desired, the lack of electric resistance in superconductors, which may be advantageous for some technologies hinders convenient voltage-bias tuning. Likewise, challenges related to nanoscale fabrication of superconductors encumber local tunability of $\psi$. Here, we demonstrate local tunability of $\psi$, obtained by patterning with a single lithography step a Nb nano superconducting quantum interference device (nano-SQUID) that is biased at its nano bridges. Our design helped us reveal also unusual electric characteristics—effective zero inductance, which is promising for quantum technologies and nanoscale magnetic sensing. Finally, we accompanied our experimental results by a semi-classical model, which not only is extending the applicability of our devices, but is also useful for describing planar nano-SQUIDs in general.

**Main Text**

The complex order parameter $(\psi = \Delta e^{i\theta})$ of mesoscopic solid-state systems plays a significant role in their functionality. Hence, controlling locally both the amplitude and phase of this order parameter is a common scientific-engineering task. In many solid-state systems, modifying $\psi$ is a challenging, or even an impossible task. For instance, in topological insulators and quantum-Hall phases, the many-body wavefunction is protected so that there is no local order parameter that can be tuned. In nano-superconducting structures, the order parameter can vary by changing material properties, such as film thickness, stoichiometry and homogeneity.[1,2] However, these characteristics are fixed for a given device and do not allow continuous tuning. Magnetic fluxes ($\Phi$) have also been used to control $\psi$. Likewise, limited control of the phase of sheet superconductors was obtained, e.g. in LaAlO$_3$/SrTiO$_3$ interfaces, while gating was demonstrated for superconductors that are integrated in ballistic semiconductors.[3–5]. In spite of these significant advances in modifying $\psi$ macroscopically, local tunability of the amplitude and phase of the complex order parameter at the device-relevant lengthscale has remained a challenge, partially, because superconductors cannot sustain electric fields.

The growing interest in topological effects in superconducting nanowires (e.g. Majoranas fermions[6–14]) as well as in the potential of such systems in realizing quantum technologies,[15–17] has raised a real necessity for tuning $\psi$ in superconducting nanowires.[18–23] Global tunability of $\psi$ is obtained when such wires are used as the weak links in superconducting quantum interference devices (SQUIDs)[20,24]. Recently, it has been shown that the magnetic flux can be introduced also more indirectly, by current-biasing the leads of a SQUID, allowing continuous control of the phase.[25–32] Yet, tunability of both the amplitude ($\Delta$) and phase ($\theta$) of the order parameter has remained elusive.[26] Moreover, local gating of $\psi$ in a small region within the wire has not been realized despite the strong technological motivation.

Here, we processed a thin Nb nano-SQUID in a single lithographic step to demonstrate continuous tunability of both the phase and amplitude of a mesoscopic superconducting order parameter. The tunability was obtained by biasing a weak-superconducting link locally. This weak link was connected in parallel with an unperturbed weak link (Fig. 1a-b) giving rise to a gated SQUID device (Fig. 1c). Consequently, the phase and amplitude of the order parameter were measured by means of the interfering critical current as a function of the applied magnetic flux. (see Fig. 1d for a complementary illustration of the interference pattern of a SQUID device). In addition to phase and amplitude tuning, we characterized the current vs. magnetic flux relation of our device. Our experimental data exhibit a strong asymmetry, which indicates an effective zero-inductance state of the device. The experimental results are accompanied by a semi-classical model, which explains both the tunability of $\psi$ and the observed asymmetry of the interference pattern.

The phase and amplitude of the order parameter are measurable by means of the interfering critical current as a function of the applied magnetic flux as follows. This periodic interference pattern arises due to the quantization of the sum of the phase differences ($\oint \nabla\theta = 2\pi m$) across the device, or

$$\sum_i \Theta_i + \frac{2\pi\Phi_{ext}}{\Phi_0} = 2\pi m, \qquad (1)$$

Here, $\Theta_i \equiv (2\pi/\Phi_0)L_i I_i + \theta_i$ are the *Fulton phases* of the $i^{th}$ branch, $L_i$ and $I_i$ are the inductance and current flow of the branch of the SQUID that contains the $i^{th}$ weak link, $\theta_i$ is the corresponding phase drops (see Fig. 2a-b for the definition of these branches), $\Phi_{ext}$ is the external flux, $\Phi_0 = h/2e$ ($h$- Plank's constant) is the quantum flux and $m$ is an integer. Electron micrographs of the gated weak link and of the SQUID are presented in Figures 2c and 2d.

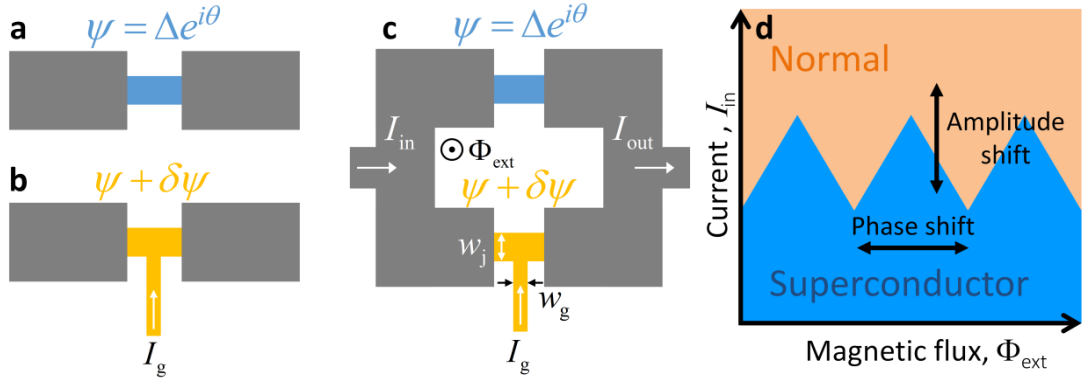

**Figure 1| Tunability of the local order parameter with a gated SQUID (schematics). (a)** Weak superconducting link embedded between two electrodes. **(b)** A gated weak link, in which the local order parameter ($\psi$) depends on the gating current ($I_g$). **(c)** The gated and the ungated weak links are connected in parallel to form a SQUID that is operated with a single input current ($I_{in}$). The output current, which is the interfering output of the device ($I_{out}$) is modulated by biasing the bottom weak link ($I_g$). **(d)** The anticipated amplitude and phase shift are observable as translations of the interfering pattern of the critical current of the entire device (i.e. the value at which the device changes from superconducting to normal) across the $I_{in}$ vs. $\Phi_{ext}$ plane. Here, phase shifts correspond to a shift along the magnetic flux axis. Likewise, amplitude shifts correspond to changes in the critical current, i.e. shifts along the input current ($I_{in}$).

According to Eq. (1), the position of the interference pattern of the device critical current ($I_c$) as a function of the external flux is determined by the local phases, $\Theta_i$. Thus, the tunability of the phase of the order parameter is detectable by the shifts of the interference pattern along the flux axis ($\delta\Theta$). Furthermore, the device critical current is determined by the critical currents of the weak links, which in turn are related to the amplitude of the local order parameter. Thus, changes in the order parameter at the weak links are observable as a shift of the interference pattern along the input-current axis (as illustrated schematically in Fig. 1d). In this work, we demonstrate the tunability of both the phase and amplitude of the local order parameter associated with a gated weak link (denoted by $i = 2$ in Fig. 2a).

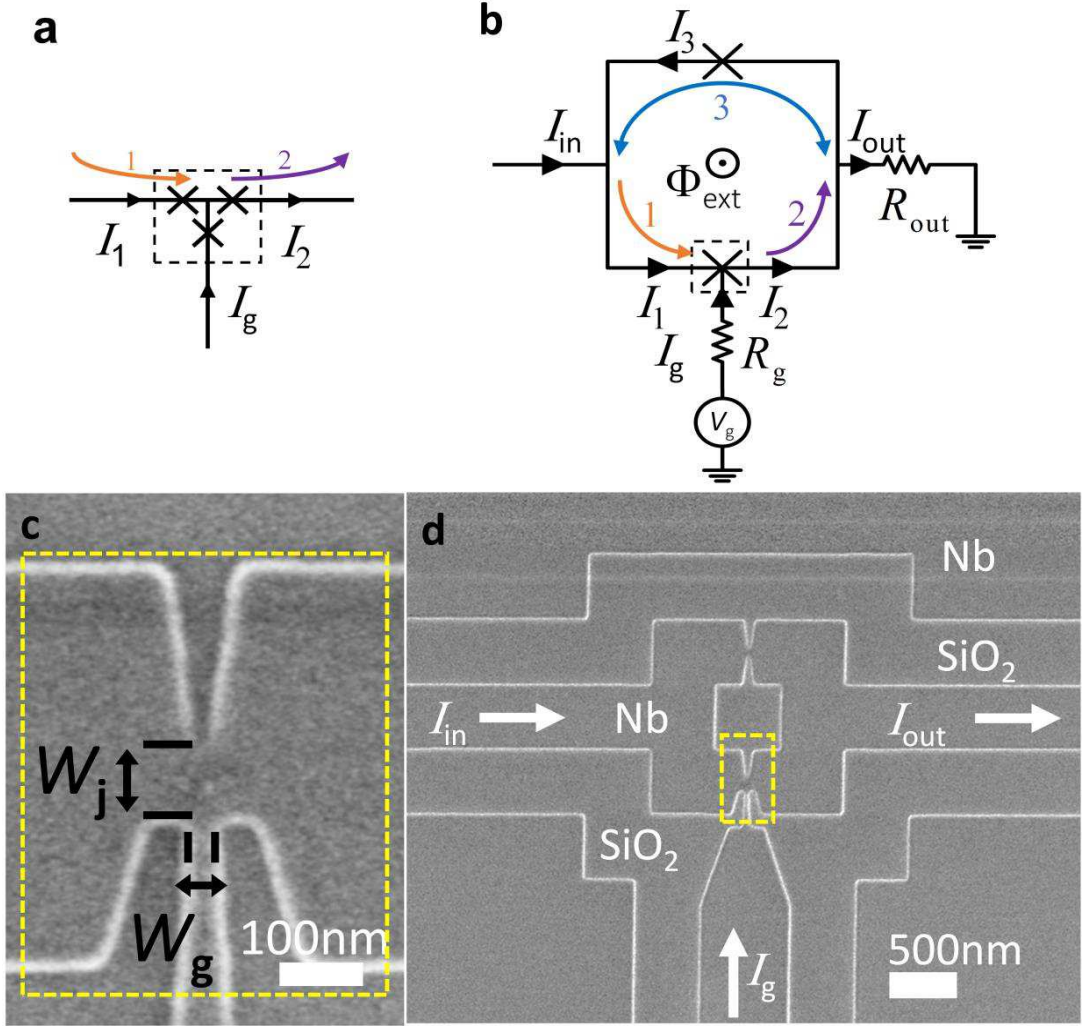

**Figure 2| Experimental setup of the gated SQUID.** (**a**) Equivalent circuit of the gated weak link, i.e. a three-port Josephson-Junction (designated by an 'x') bundle. (**b**) Equivalent circuit of the gated SQUID (the gated weak link is designated by a dashed square with an 'x' inside). (**c**) SEM image of our Nb on SiO$_2$ gated weak-link ($w_g$ denotes the width of the gate port, and $w_j$ denotes the width of the weak-link port). (**d**) SEM image of our gated DC-SQUID. The input- ($I_{in}$), gating- ($I_g$) and output- ($I_{out}$) currents flow in the device are added to guide the eye.

To examine the tunability of the local complex order parameter we modulated the external parameters, the applied current $I_{in}$, external magnetic flux $\Phi_{ext}$ and gate voltage $V_g$ (interchangeable with the gate current, $I_g$), while measuring the voltage dropped on our device (electric schematic is shown in Fig. 2b). The fabrication process is detailed in Supplemental Information. The fabricated devices differed by the ratio between the width of the gating port and the weak-link port $w_g/w_j$ (corresponding to Fig. 2c). The leads were wire-bonded to a chip carrier and the entire device was inserted to a 3.2 K cryostat. The SQUID was then tested electrically, by applying a DC current (practically, a slow changing signal at 1 kHz to avoid latching) and introducing a magnetic flux through a superconducting coil (see Supplemental Information). The periodic interference pattern of the SQUID that is given in Fig. 3a reveals a zigzag interference shape. In total, we tested 24 devices from 5 chips, all showed a similar behavior to the results discussed below.

We modelled our gated weak-link as three separated Josephson junctions (JJs) as shown in Fig. 2a. Following the notation in Fig. 2a-b, we took a closer look at the influence of the gating on the order parameter at the branch *i*=2, i.e. $\psi_2$. To demonstrate controllable phase shift, we modulated the gate current $(I_g)$, while keeping the other parameters unchanged (practically, we applied a voltage, $V_g$ across a resistor, $R_g$). As a result, we were able to change the periodic interference pattern along the flux axis, demonstrating controllability of Θ with the gating current as shown in Fig. 3a. For large gate currents, the interference pattern disappeared due to the crossing of the critical current of the gate port and thus limiting the phase shift to half a period. However, by introducing a device in which the two weak links were gated (Figures SI2a and SI2b), we obtained complete control over the phase (Fig. SI2c). Due to the device asymmetry, only positive applied currents ($I_{in} > 0$) and positive gate voltage ($V_g > 0$) resulted in a zigzag interference pattern, while positive applied currents ($I_{in} > 0$) with negative gate voltages ($V_g < 0$) did not show the zigzag pattern (shown in Fig. SI1a), due to the crossing of the critical current of the gate port. Finally, inverting both the applied current ($I_{in} < 0$) and the gate voltage ($V_g < 0$) resulted again in the zigzag interference pattern, as expected (shown in Fig. SI1b).

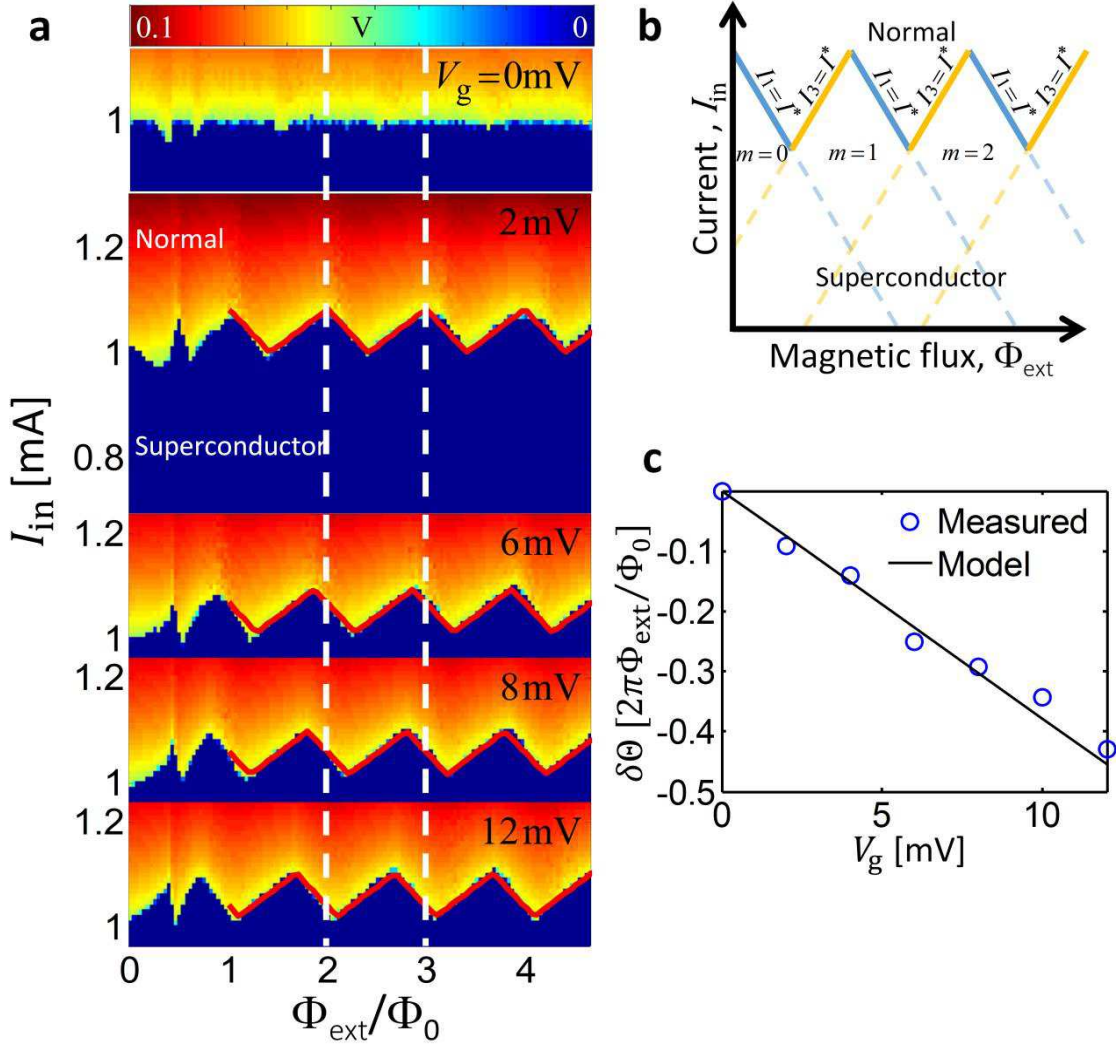

**Figure 3| Phase tunability in a gated SQUID**. (**a**) Experimental results of the interfering output (device resistance, following the top colormap) from a gated SQUID for chosen gating voltage ($V_g$) values. A phase shift in the interference pattern is observed as a function of the gate voltage (motion along the external flux, $\Phi_{ext}/\Phi_0$ axis). The white dashed lines help trace the phase shift. Red lines are the best-fit prediction of the critical current of the device, by using the presented semiclassical model, demonstrating a good agreement with the experimental results. (**b**) Schematic illustration of the presented model, in which the interfering pattern is a result of a crossover between two sets of parallel lines, which are the solutions of the internal critical currents $I_1$ and $I_3$ (following Fig. 2a-b) in Equations (SI13)-(SI15). These lines are a linear function of external device parameters: input current $I_{in}$, external magnetic field $\Phi_{ext}$, and the number of fluxons $m$, while their translation along the external-flux axis (i.e. phase shift $\delta\Theta_2$) is set by the gate current (see SI for more details). (**c**) The measured phase shift ($\delta\Theta$) as a function of applied gate voltage (blue circles) as extracted from (a), while the calculated dependence of the phase shift ($\delta\Theta_2$) on the applied voltage from our model (Eq. (3)) is given by the black line.

The large number of the local JJs of interest in our device (four for a single gated weak-link device) and the circuit geometry (Fig. 2a-b) encumber the extraction of an analytical expression[33] for the dependence of the measured critical current of the entire device ($I_c$) on $\Phi_{ext}$. Hence, we introduced a semiclassical model that complements the quantization condition of Eq. (1) with the appropriate Kirchhoff's current and voltage laws (following the schematics in Fig. 2b). First, we used this model to define our SQUID operation framework, i.e. the zigzag interference pattern in Fig. 3a. Following Equations (SI1)-(SI5) we approximated the total Fulton phase by:

$$\Theta \approx \frac{2\pi}{\Phi_0} \sum_i L_i I_i + \Theta_0 \quad (2)$$

where $\Theta_0$ is a constant that depends on the device geometry and $I_i$ ($i = 1,2,3$) is the current flows in one of the three branches of the SQUID (denoted in Fig. 2a-b). As detailed in the SI (Equations (SI3)-(SI4)), the *linearized* quantization condition of Eq. (2) is useful for modeling the critical currents of SQUID devices with large $\beta_{L,i}$'s ($\beta_{L,i} \equiv \left(\frac{2\pi}{\Phi_0}\right) L_i I_i^* \gg 1$, where $I_i^*$ is the critical current in the $i^{\text{th}}$ branch). Using this condition, we can express the internal currents that flow in the three branches of the SQUID ($I_i$'s) as a *linear* function of the external device parameters: input current $I_{\text{in}}$, gate current $I_g$, external magnetic field $\Phi_{\text{ext}}$ and the number of fluxons, $m$ (Equations (SI13)-(SI15)). Using these expressions of the $I_i$'s, we extracted the critical current of the entire device, $I_c$, by demanding that for $I_{\text{in}} < I_c$, all three internal currents are below their critical values ($I_i < I_i^*$). The zigzag interference pattern of $I_c$ vs. $\Phi_{\text{ext}}$ is then determined by the intersections between the linear solutions of the three conditions for the critical currents: $I_1 = I_1^*$, $I_2 = I_2^*$ and $I_3 = I_3^*$. These intersections are illustrated graphically in Fig. 3b for the specific case of $I_1^* = I_2^* = I_3^*$ and $I_1 > I_2$. Fig. 3c shows a good qualitative agreement of our model to the experimental results. The detailed derivation of this fit is given in the SI, where we demonstrate that for an ungated SQUID device, our approach coincides with the known large-inductance approximation ($\beta_L \to \infty$).[33] Finally, our linear simplification allows us to extend the model to circuits with a large number of weak links.

This model naturally accounts for the observed horizontal shift of the interference pattern induced by the gate voltage. For a fixed value of the critical input current, both $I_1$ and $I_3$ are unchanged (one of them equals to $I^*$ and their sum equals to $I_{\text{in}}$), and $\delta I_2 = \delta I_g$. Within the framework of the present linearized quantization condition the desired phase shift tunability as a function of gating current is then given by:

$$\delta\Theta = -\frac{2\pi}{\Phi_0} L_2 \frac{V_g}{R_g + R_{\text{out}}}. \quad (3)$$

This phase shift is entirely attributed to a phase shift of the second branch $\delta\Theta_2 = (2\pi/\Phi_0) L_2 \delta I_2$ (detailed analysis is shown in SI-3 section in the SI).

After having controlled the superconducting phase ($\Theta_2$) experimentally, we demonstrated that tuning the critical current ($I_2^*$) is also possible (Fig. 4a-f). Here, we used the same experimental circuit (Fig. 2a-b), but only now the gate terminal was narrower than the weak-link width, confining the channel even further ($w_g/w_j < 1$). For low gating currents, the device showed a bistable behavior, with a non-monotonous resistance as a function of $I_{\text{in}}$ (Fig. 4c). However, further increasing $V_g$ changed the measured interference pattern in three aspects as demonstrated in Fig. 4a-f: (i) increasing the gate voltage (Fig. 4e-f, for which $V_g > 10.8$ mV) distorted the periodic envelope, so that a strong asymmetry was induced, leading eventually to a region in which $dI/d\Phi \to \infty$ (highlighted in Fig. 4e); (ii) the device *critical current*, which is the 'envelope' of the periodic interference pattern (see Fig. 1d) moved towards lower-current values as demonstrated in Fig. 4d-f; and lastly, at even-higher gating voltages, (iii) the modulation of the envelope (i.e. peak to peak value) decreased (Fig. 4f). The reduction of the critical current of the entire device is shown in Fig. 4g.

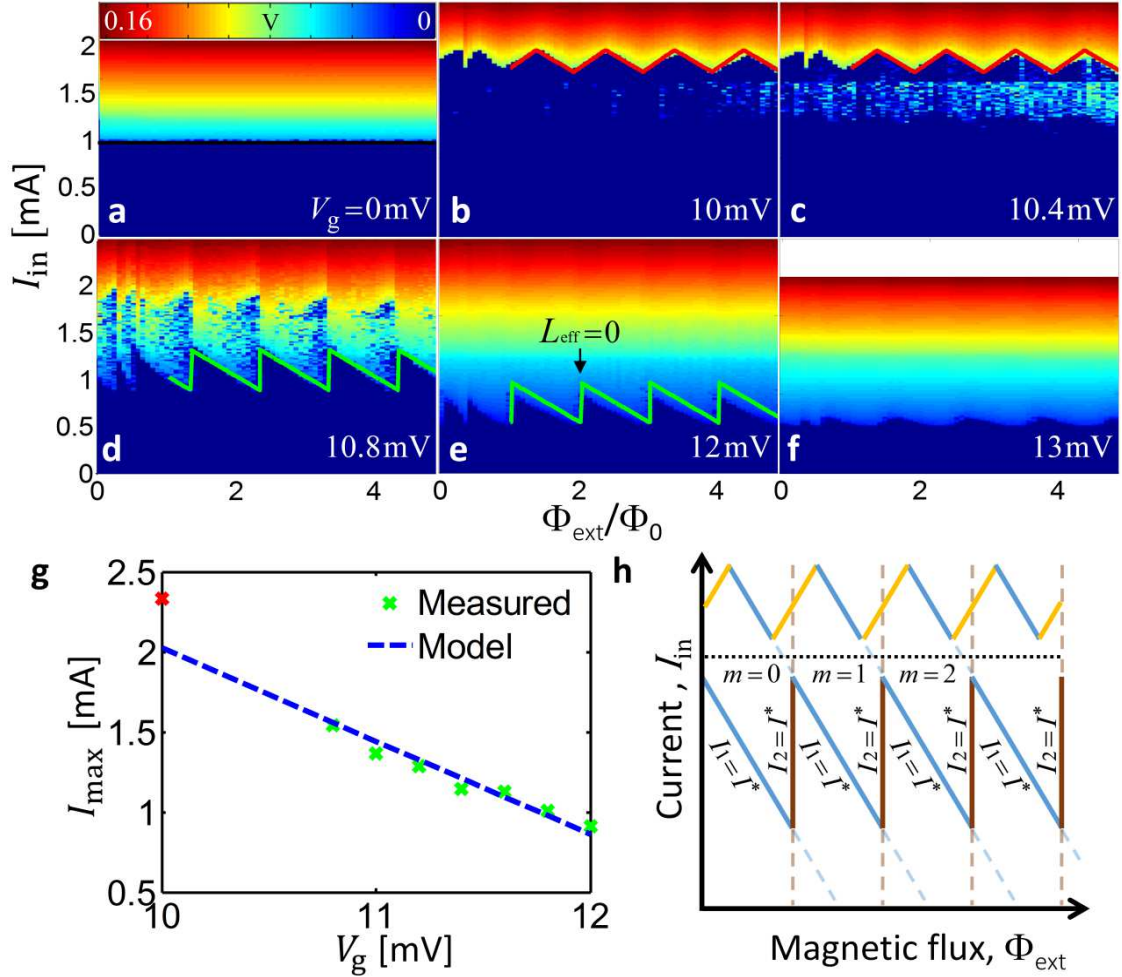

**Figure 4| Amplitude tunability in a gated SQUID**. (**a-f**) Experimental results of the interfering currents (following the top colormap) from a gated SQUID for chosen gating voltage $(V_g)$ values. (**d-f**) An amplitude shift in the interference pattern is observed as a function of the gate voltage $V_g$ (motion along current axis, $I_{in}$). Here, $w_g/w_j < 1$. In (a), for $V_g = 0$ mV no interference pattern is observed due to $I_g$ exceeding the critical current of the gate port $(I_g^*)$. (b) For $V_g = 10$ mV, the SQUID interference pattern is observed which corresponds to the interference pattern between two branches of the SQUID $I_1$ and $I_3$. The red line shows the semiclassical model calculation superimposed on the experimental data. (c) For $V_g = 10.4$ mV we see a metastable region in addition to the SQUID interference pattern. (d) For $V_g = 10.8$ mV the interference pattern is distorted and strong asymmetry is induced (the green line correspondences to the semiclassical calculation superimposed on the experimental data). Note that the interference pattern changes its form to a region with $dI/d\Phi \to \infty$ which can be interpreted as $L_{eff} \to 0$. Moreover, reduced critical current of the device is observed which corresponds to the "envelope" of the interference moving towards lower-current values as shown in (e). (f) For higher gating $V_g = 13$ mV, the modulation of the envelope (i.e. peak to peak) is decreased. (**g**) Measured critical current of the entire device as a function of gate voltage $V_g$ with the semiclassical model as described by Eq. (4). Colors of the measured symbols follow the colors in the fitting curves in (a-f). (**h**) Shows solution of the semiclassical model with the linearized quantization condition using the schematic of Fig. 2a and 2b, which results in an interference pattern as was measured in (a-f). The approximation is detailed in the SI (see Equations. (SI17)-(SI18)) using the effective description of the coupled junctions as $I_1^* = I_2^* = I^* - \alpha |I_g|$.

To accommodate our model for the devices with narrow gates, we first accounted for the measured strong asymmetry of the zigzag pattern (see Fig 4e at $V_g = 12$ mV). Here, we assumed that the critical currents of the biased weak links ($I_1^*$ and $I_2^*$) were suppressed by the gate current, while the unbiased weak link ($I_3^*$) remained unchanged. A faithful theoretical description of this effect should involve coupling as well as energy transfer between three strongly coupled nonlinear oscillators (i.e. the weak links located in proximity to the gate) and is hence extremely challenging. We can nevertheless express the influence of the gate voltage $V_g$ on $I_1^*$ and $I_2^*$ within the present *linear* effective description: $I_1^* = I_2^* = I^* - \alpha |I_g|$, where $\alpha > 0$ represents the coupling strength between the weak links, and $I_g = (V_g - R_{out}I_{in})/(R_g + R_{out})$ is the current flowing through the gate. As shown in the SI, this effect leads to a shift of the zigzag pattern along the current axis (corresponding to an amplitude change) by:

$$\delta I_c = \frac{V_g}{R_g}\frac{1+\alpha}{1-\alpha\, r}, \quad (4)$$

where $r = R_{out}/R_g$ is the ratio between the gate and output resistances. Fig. 4g shows Eq. (4) compared with the measured maximal critical current of the device.

Our experimental measurements show a sharp jump of the critical current as a function of the magnetic flux, or $dI_c/d\Phi_{ext} \to \infty$ (denoted in Fig 4e). This sharp jump may hold important technological consequences because it implies that our device is extremely sensitive to small variations of the magnetic field. Equivalently important, following Faraday's law of inductance, this sharp jump indicates that our device demonstrates effectively zero inductance $L_{eff} = (dI_c/d\Phi_{ext}))^{-1} \to 0$. The absence of inductance, is not only interesting scientifically, but it is also technologically attractive. We should note that the lack of inductance occurs exactly where $m$ changes (i.e. where there is a sudden change in one quantum fluxon in the system). This behavior is not observed in typical SQUIDs, where the two branches play equivalent roles and do not lead usually to an asymmetric interference pattern.

The coupling parameter that describes the strong asymmetry of our system ($\alpha$) is estimated from the experimental observations as follows. The limit $dI_c/d\Phi_{ext} \to \infty$ is achieved when the equation $L_3 - (L_1 + \alpha \sum_i L_i)r = 0$ is satisfied. Substituting the experimental parameters ($L_1 = L_2 \approx L_3/2$, $r = 0.57$) results in $\alpha \approx 0.8$. This large value of the coupling constant indicates that introducing the gating bias in the case of a narrow gate terminal reduced significantly the critical current of the individual neighboring weak links ($I_1^*$ and $I_2^*$) and consequently the critical current of the entire device ($I_c$). The interplay between these two effects is further highlighted by Eq. (4), which shows that for $\alpha < 1/r$, the offset of the critical current of the device grows with increasing $\alpha$. For $\alpha > 0.8$, our model predicts an unphysical re-entrant behavior of the interference pattern. This scenario may explain the observed bistable behavior in the case of intermediate bias values (see Fig. 4c). Qualitative agreement of our model with the experimental data is illustrated in Fig. 4h, while quantitative fitting of our experimental data to the model is given by the solid lines at the superconducting-normal transition in Fig. 4a-f (in the black, red, and green lines superimposed on the measured data).

The strong coupling between the weak links deserves further investigations and raises fundamental questions concerning coupling mechanisms in nano-superconducting junctions. For instance, our results suggest that gated devices with tunable complex order parameter are useful for realizing local tunability of the superconducting-to-insulating transition in nanostructures. Lastly, our work offers another angle to the increasing interest in the study of multi-terminal Josephson junctions (see the recent work by Padurariu et al[34] and references therein). Specifically, our demonstrated method for tuning electrically the local quantum, or mesoscopic state of a nano three-terminal device may pave the way to novel quantum and energy-efficient switching technologies.

**Methods**

The device processing and characterization methods as well as the complete derivation of the presented model are given in the Supporting Information.

**Acknowledgments**

The authors would like to acknowledge financial support from Nevet. YI and IH thank the Horev Fellowship for Leadership in Science and Technology, supported by the Taub Foundation and the Zuckerman STEM Leadership Program. EB thanks the Israeli Science Foundation. EGDT is supported by the Israel Science Foundation, grant N. 1452/14, while RW would like to thank RBNI for the nanofabrication facilities and financial assistance.

# Supporting Information

**Local tuning of the order parameter in superconducting weak links: A zero-inductance device**

**Contents**

**SI-1. Experimental Method and Material**

In this section we provide details about the fabrication process. The fabrication process of our devices was as follows the framework of the linear quantization. First, a 35-nm-thick Nb film was deposited (by means of DC magnetron sputtering) on top of an intrinsic Si substrate with a native-oxide top surface, followed by spin coating of PMMA 950 A3 at 3500 rpm. Next, the pattern was formed by using electron-beam lithography (EBL). Finally, the SQUID was fabricated by using reactive ion etching (RIE) with a $CF_4$ gas. The fabricated SQUID was glued to a chip holder and bonded using aluminum wires. To measure the SQUID, we used a closed-cycle cryogenic system with DC wiring. The measurements were done with a low frequency AC sine function generator at 1 kHz (Agilent 33220A). A 1 kΩ resistor was connected in series to the device for the input current of the SQUID, $I_{in}$. The gate port of the SQUID was connected to a voltage supply (Yokogawa 7651 programmable dc source). Magnetic field was tuned using an Oxford Instruments IPS-120 power supply and a superconducting magnetic coil.

**SI-2. Device operation asymmetry**

In Fig. SI1 we presented additional operation regimes as described in the main text (in the discussion related to Fig. 3). In Fig. SI1a we showed positive input currents ($I_{in} > 0$) with negative gate voltages ($V_g < 0$), while in Fig. SI1b we showed negative input current ($I_{in} < 0$) with negative applied gate voltage ($V_g < 0$). As can be seen in Fig. SI1a, the interference pattern is not present in the $I_{in} > 0$, $V_g < 0$ scenario, owing to the crossing of the gate port critical current, which is lower than the critical currents in the SQUID. For symmetric operation (as shown in Fig. SI1b) the zigzag interference pattern was observed as expected.

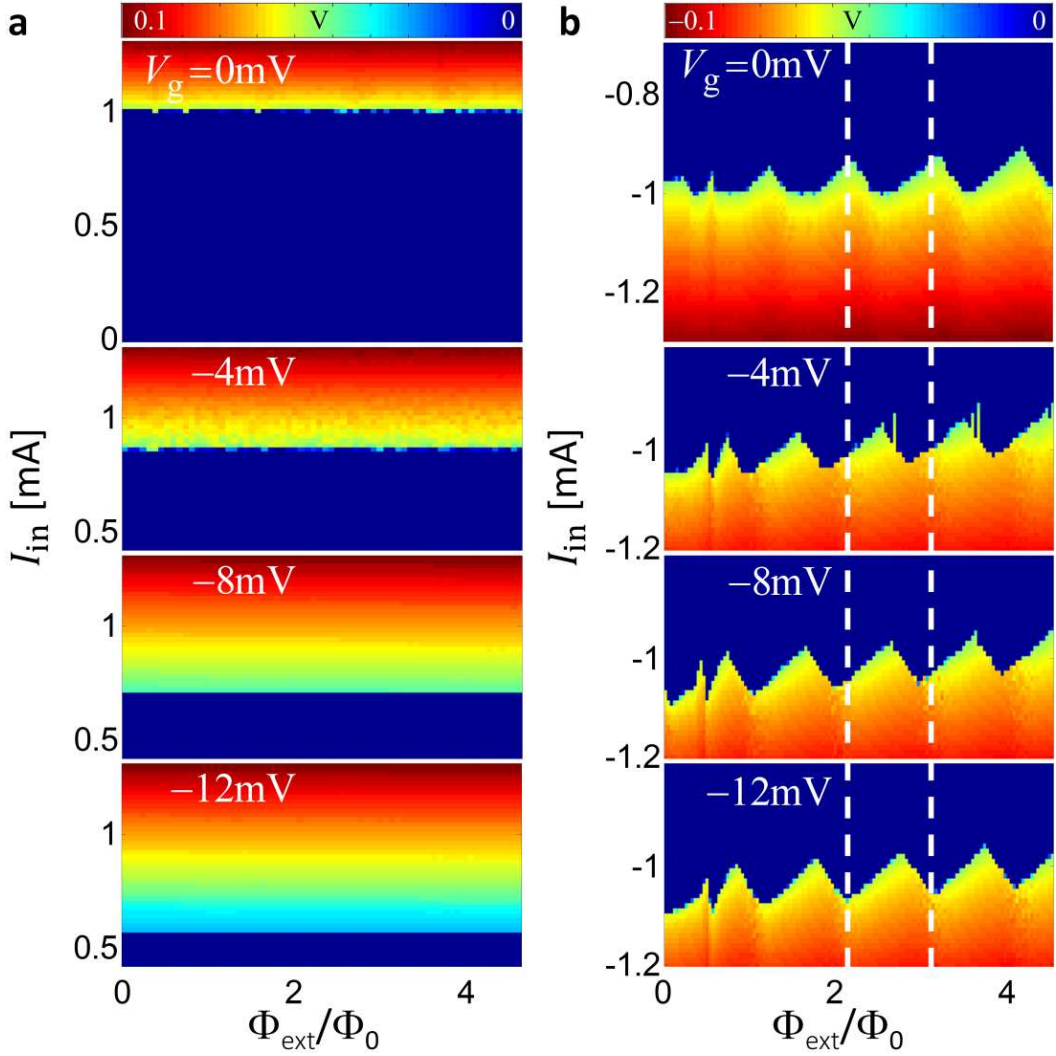

**Figure SI1 Gated SQUID under varying voltage and current polarities**. (**a**) Experimental measurements of the SQUID device using positive applied currents ($I_{in} > 0$) with negative gate voltages ($V_g < 0$), which results with no interference pattern. (**b**) Measurements of the SQUID device using negative applied currents ($I_{in} < 0$) with negative gate voltages ($V_g < 0$), which results with an interference pattern.

We note that in the magnetic flux regime there is a sharp dip in the first oscillation (visible in Figures 3a, 4a-f, and SI1). These artifacts were also observed for simple DC-SQUIDs (no gating ports) measured in our system. Thus, these dips are unlikely to be associated with the gated SQUID design. The origin of these artifacts is probability due to the use of thin superconducting films (~35nm), thinner than the London penetration depth of Nb ($\Lambda \approx 40$ nm). That is, these dips may originate from flux penetration. However, the exact origin of these dips is not yet completely clear to us.

**SI-2. Double gated SQUID design**

To achieve complete control over the order-parameter phase we fabricated a double gated SQUD device as illustrated in Fig. SI2a. SEM image of our double gated SQUID design is shown in Fig. SI2b. By using two sources for each port (as shown in Fig. SI2a), we measured the interference pattern and calculated the phase shift resulted from the applied voltage on the gate ports (Eq. 3). In Fig. SI2c, the phase shift for each is shown as a function of the gate voltage $V_{g1}$ and $V_{g2}$, demonstrating complete tunability of the phase.

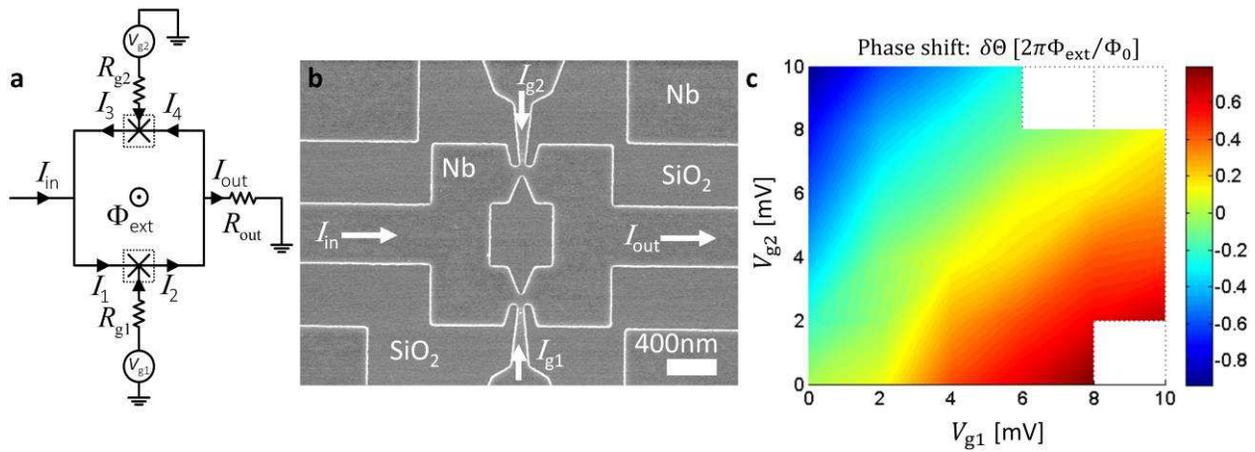

**Figure SI2 Complete tunability of the phase in a double-gated SQUID.** (**a**) Equivalent circuit of the double gated SQUID design (the two gated weak links are designated by a dashed square with an 'x' inside). (**b**) SEM image of our double gated DC-SQUID. The input- ($I_{in}$), gating current- ($I_{g1}$) $and$ ($I_{g2}$), and output current ($I_{in}$) are shown in white. (**c**) The measured phase shift ($\delta\Theta$) (shown in color) as a function of applied both gate voltages $V_{g1}$ and $V_{g2}$ as extracted from the data.

**SI-3. Linearized quantization framework for SQUID device**

In the following sections (SI-3a and SI-3b) we discuss the framework of the linear quantization conditions as well as the derivation of linearized quantization condition that is presented in the paper (Eq. (2) of the main text). Furthermore, we discuss the applicability of this model and generalize it to any DC-SQUID with large inductance ($\beta_L \gg 1$). Finally, we compare our semiclassical model with an analytical model of a simple DC-SQUID (without a gate port) and we discuss its approximation validity. This derivation demonstrates the advantageous applicability of the model to multi-port or multi-junction devices that are difficult to model otherwise.

In section S-3b we use the linearized quantization approach to derive analytical expressions for the phase shift and critical current of our gated device. This approach is used to obtain the theoretical interference patterns shown in Figs. 3b and 4h of the main text as well as to obtain equations (3) and (4).

### SI-3a. Derivation of the linearized quantization condition

Our starting point is the quantization of the superconducting phase over a closed path. We can now apply an external magnetic flux and account for the circulating current, $\Phi_T = \Phi_s + \Phi_{ext} = \sum_i L_i I_i + \Phi_{ext}$, so that:

$$\frac{2\pi}{\Phi_0}\left[\sum_i L_i I_i + \Phi_{ext}\right] + \sum_i \theta_i = 2\pi m. \tag{SI1}$$

Here $i$ represents the segments of the closed path (Fig. 2a-b), $L_i$ is the total inductance across the segment $i$ (given by the sum of the kinetic and geometric components), $\Phi_{ext}$ is the external magnetic flux, $\theta_i$ is the phase drop across the weak link in the segment $i$ (Fig. 2a-b), and $m$ is the number of fluxons. We now introduce the Fulton phase[1,2], $\Theta_i = \theta_i + \frac{2\pi}{\Phi_0} L_i I_i$ (note that $\theta_i$ depends on $I_i$ from the current-phase relation (CPR)), and we can express the quantization condition of the device:

$$\sum_i \Theta_i + \frac{2\pi \Phi_{ext}}{\Phi_0} = 2\pi m. \tag{SI2}$$

We can now define the Fulton phase of the entire device as $\Theta \equiv \sum_i \Theta_i$, and the quantization condition becomes:

$$\Theta + \frac{2\pi \Phi_{ext}}{\Phi_0} = 2\pi m.$$

To fully define the problem and to apply the linear semiclassical approximation model, we need to supplement Eq. (SI2) with the CPRs of the weak links. For simplicity we assume the Josephson relation $I_i(\theta_i) = I_i^* \sin(\theta_i)$, where $I_i^*$ is the critical current of the $i^{th}$ weak link (generalization to non-sinusoidal CPR will be discussed later).

Let us now consider a situation relevant to the calculation of the critical current of a DC-SQUID with large inductance ($\beta_L \gg 1$). As we will self-consistently check below, in such devices the variations of the critical current are small. Thus, at the threshold condition, all the internal weak links are operated close to their critical current, i.e. $I_i^* - I_i \ll I_i^*$. Having defined $\delta\theta_i = \frac{\pi}{2} - |\theta_i|$ and $I_i = I_i^* - I_i$, we can approximate the CPR by Taylor expansion: $\delta I_i \approx \frac{1}{2} I_i^* \delta\theta_i^2$, or $\delta\theta_i = \sqrt{2\delta I_i/I_i^*}$. Thus, the Fulton phase of each junction becomes

$$,\Theta_i \approx \frac{\pi}{2} - \sqrt{\frac{2\delta I_i}{I_i^*}} + \frac{2\pi L_i}{\Phi_0}(I_i^* - \delta I_i) = \frac{\pi}{2} + \frac{2\pi L_i I_i^*}{\Phi_0} - \frac{2\pi L_i \delta I_i}{\Phi_0}\left(1 + \sqrt{\frac{2 I_i^*}{\beta_{L,i}^2 \delta I_i}}\right) \quad (SI3)$$

where $\beta_{L,i} = \left(\frac{2\pi}{\Phi_0}\right) L_i I_i^*$. When $\delta I_i/I_i^* \gg 1/\beta_{L,i}^2$ Eq. (SI3) yields

$$\Theta_i \approx \Theta_i^* + \frac{2\pi L_i I_i}{\Phi_0}, \quad (SI4)$$

where we defined $\Theta_i^* = \left(\frac{\pi}{2}\right)\text{sign}(I_i)$. Note that Eq. (SI4) is trivially satisfied also in the case $\delta I_i = 0$. We can now substitute Eq. (SI4) into (SI2) and obtain

$$\frac{2\pi}{\Phi_0}\left(\sum_i L_i I_i + \Phi_{ext}\right) + \Theta_0 = 2\pi m, \quad (SI5)$$

where $\Theta_0 = \sum_i \Theta_i^*$ is equivalent to a constant offset in the magnetic flux. Note that $\Theta_0 = 0$ in symmetric devices (i.e. where the number of weak links in the upper and lower branches is equal). Eq. (SI5), which is perhaps the main outcome of our model, offers a quantization condition that is linear in the internal currents $I_i$'s. This equation allows us to obtain analytic expressions for the critical current of the device, as a function of the external currents, the magnetic flux, and the number of fluxons. The simplification involved with this equation with respect to other analytical derivations of the CPR in DC-SQUIDs allows us to solve analytically the current distribution also in complicated DC-SQUIDs[3–5] circuits, which is difficult to derive otherwise, as in the case of the present work.

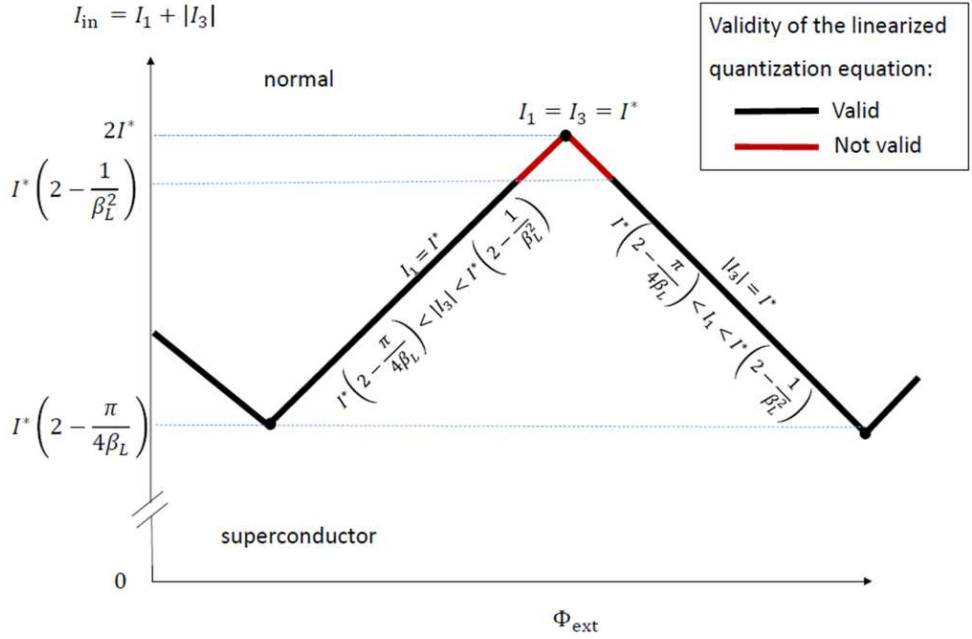

**Figure SI3 The framework of our model**. Validity of the linearized quantization condition to describe the superconducting-normal transition of SQUID devices with large inductance. The plot refers to the simplest case where $I_i^* = I^*, L_i = L, I_g = 0$ s.t. $I_1 = I_2$.

Let us now discuss the validity of Eq. (SI5) and its applicability to the computation of the critical current of the device. In above derivation, the first approximation, (SI3), relied on the assumption that $\delta I_i / I_i^* \ll 1$. In a simple (normal) DC-SQUID the slope of the zigzag pattern is simply given by $1/L_{\text{tot}} \equiv (\sum_i L_i)^{-1}$. In this case, the internal current varies by $\delta I_i \sim \Phi_0 / L_{\text{tot}} \sim (I_i^* / \beta_L)$ near the edges of the zigzag. Thus, our approach is valid as long as $\beta_L \gg 1$. A similar argument holds also in the low-bias part of Fig. 4a-e (main text). However, for large biases (Fig. 4f), where $\delta I_i$ becomes of the order of $I_i^*$ these assumptions are not valid any more. The second approximation, Eq. (SI4), is valid if all the internal currents $I_i$'s satisfy $1/\beta_{L,i}^2 \ll 1 - I_i/I_i^* \ll 1$, or equals to $I_i^*$. This condition generically applies to the superconducting-normal transition of the device, where one of the weak links is precisely at its critical current, and the others oscillate between $\approx I_i^*(1 - 1/\beta_{L,i})$ and $I_i^*$. Note that for large inductances $1/\beta_{L,i}^2 \ll 1/\beta_{L,i}$, and thus the linearized condition is valid for most of the zigzag pattern (see Fig. SI3). This analysis suggests that Eq. (SI5) is valid in the limit of $\beta_{L,i} \to \infty$. Indeed, in the case of a simple DC-SQUID (made by two parallel Josephson junctions), the linear quantization condition (SI5) coincides with the analytic results of Segev et al.[3] for this limit (as shown in Fig SI4, which compares the exact solution of the non-linear analytical model[3–5] and the approximated model). For finite $\beta_{L,i}$ we expect Eq. (SI5) to correctly capture the position of the critical current and its dependence on the external parameters, although the exact shape differs from the predicted straight lines (as demonstrated in Fig. SI4). Finally, we should note that our derivation is valid for any general CPR that is described by an analytic function (which satisfies $dI_i/d\theta_i = 0$), and our model thus is not restricted to ideal Josephson junctions. We therefore believe that using our model for devices that satisfy this general condition can help analyze their behavior significantly.

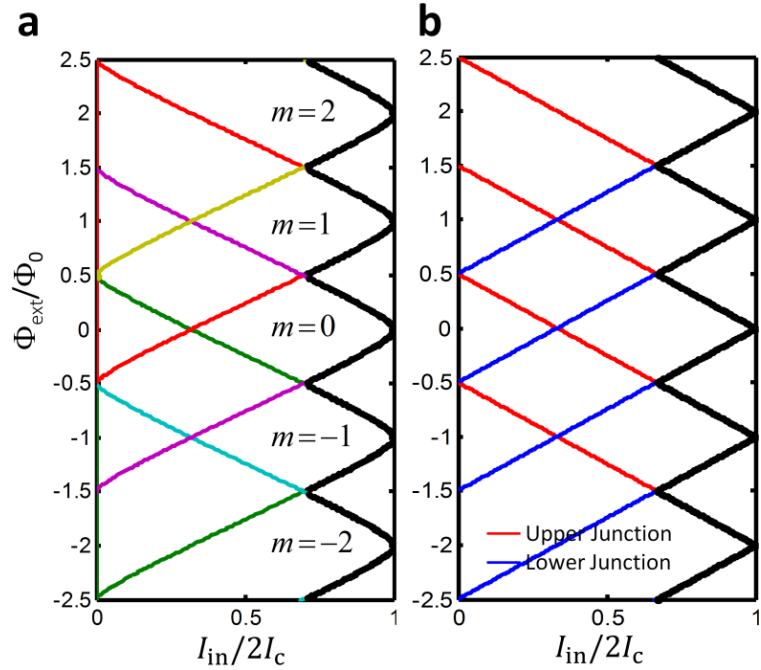

**Figure SI4 Comparison between our model and the exact circuit analysis** (**a**) The stability regions as calculated by the full stability analysis[3–5] using $\beta_L = 2$ demonstrates the applicability of the semiclassical model. Black line represents the crossover between the superconducting and normal state. Colored lines represent a stability region for each quantum integer $m$. (**b**) Solution of the semiclassical model of a simple DC-SQUID, where the blue lines represent the lower junction being triggered, and the red lines represent the upper junction being triggered (upper and lower refer to Fig. 2b of the main text). Each blue (red) line corresponds to different flux quantum $m$ as in (a).

## SI-3b. Semiclassical model of a gated SQUID

After deriving the general model in Section SI-3a and discussing broadly its framework, in this Section we demonstrate how it was implemented to model our gated devices. Our starting point is Kirchhoff's current and voltage laws of the circuit presented in Fig. SI5:

$$I_{in} = I_1 - I_3, \ I_g = -I_1 + I_2, \ I_{out} = I_2 - I_3, \ I_{out} = V_0/R_{out}, \ I_g = \frac{-V_0+V_g}{R_g}. \tag{SI6}$$

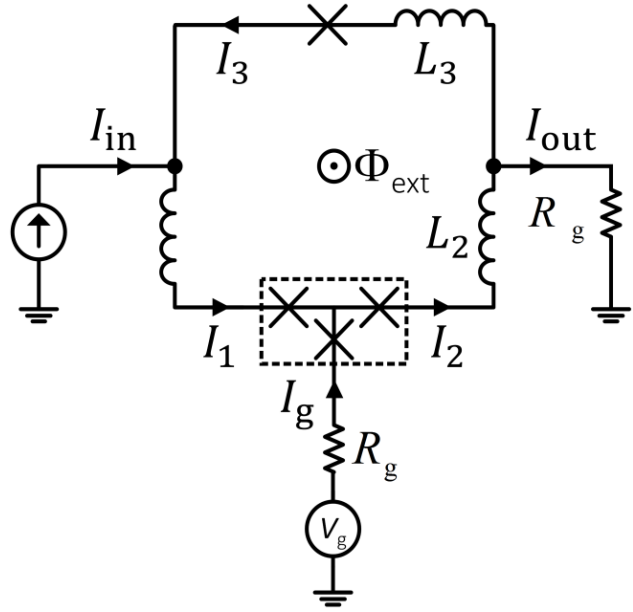

**Figure SI5** Detailed electrical effective schematics model of the device.

From these equations, we obtain the relation between the gate current and the gate voltage:

$$I_g = \frac{V_g - R_{out}I_{in}}{R_g + R_{out}}. \tag{SI7}$$

Because the gate current is introduced in the device through a weak link, the absolute value of $I_g$ must not exceed a threshold value, which we denote by $I_g^*$. This condition ($|I_g| < I_g^*$) imposes an upper bound on the input current $I_{in} < I_c^{(g)}$, where

$$I_c^{(g)} = \frac{V_g - I_g^* R_{out}}{R_g + R_{out}}. \tag{SI8}$$

In our devices, the interference pattern disappears for $V_g = 0$, where $I_c^{(g)}$ becomes smaller than the critical current induced by the internal weak links. When $V_g$ is increased above a critical threshold, the critical current of the device becomes dominated by the internal weak links and the interference pattern reappears.

To compute the internal currents of the device, we combined equations (SI6) with the linearized quantization condition (equations (1) and (2) of the main text), to obtain

$$\frac{2\pi}{\Phi_0}(L_1 I_1 + L_2 I_2 + L_3 I_3 + \Phi_{\text{ext}}) = 2\pi m + \Theta_0. \tag{SI9}$$

By combining the quantization condition SI9 with the Kirchhoff laws SI6, we expressed the internal currents of the device as a function of the external sources $I_{\text{in}}, V_g$, resistances $R_g$ and $R_{\text{out}}$, and inductances $L_1, L_2$ from Eq. (SI6) and Eq. (SI9)

$$I_1 = \frac{L_2(-V_g + R_{\text{out}} I_{\text{in}}) + (R_g + R_{\text{out}})\left(L_3 I_{\text{in}} + \left(m + \frac{\Theta_0}{2\pi}\right)\Phi_0 + \Phi_{\text{ext}}\right)}{(L_1 + L_2 + L_3)(R_g + R_{\text{out}})}, \tag{SI10}$$

$$I_2 = \frac{L_3(V_g + R_g I_{\text{in}}) + L_1(V_g - R_{\text{out}} I_{\text{in}}) + (R_g + R_{\text{out}})\left(\left(m + \frac{\Theta_0}{2\pi}\right)\Phi_0 + \Phi_{\text{ext}}\right)}{(L_1 + L_2 + L_3)(R_g + R_{\text{out}})}, \tag{SI11}$$

$$I_3 = -\frac{(L_2(V_g + R_g I_{\text{in}}) + (R_g + R_{\text{out}})\left(L_1 I_{\text{in}} - \left(m + \frac{\Theta_0}{2\pi}\right)\Phi_0 - \Phi_{\text{ext}}\right)}{(L_1 + L_2 + L_3)(R_g + R_{\text{out}})}. \tag{SI12}$$

In the framework of our experimental results, the current of interest is the critical current of the entire device, $I_c$. We follow a semiclassical approach to relate this quantity to the internal currents ($I_i$'s) and assume that the device becomes normal as soon as one of the three branches reaches a threshold value $I_i^*$. By using equations (SI10-SI12) and demanding $I_i = I^*$, we obtain three critical conditions for the input current, (denoted by $I_c^{(i)}$):

$$I_c^{(1)} = \frac{-(R_g + R_{\text{out}})\left(m + \left(m + \frac{\Theta_0}{2\pi}\right)\Phi_0 + \Phi_{\text{ext}} - (L_1 + L_3)I^*\right) + L_2(V_g + (R_g + R_{\text{out}})I^*)}{L_2 R_{\text{out}} + L_3(R_g + R_{\text{out}})}, \tag{SI13}$$

$$I_c^{(2)} = \frac{-(L_1+L_3)V_g-(R_g+R_{out})\left(\left(m+\frac{\Phi_0}{2\pi}\right)\Phi_0+\Phi_{ext}\right)+(L_1+L_2+L_3)(R_g+R_{out})I^*}{L_3R_g-L_1R_{out}}, \tag{SI14}$$

$$I_c^{(3)} = \frac{(R_g+R_{out})\left(m+\left(m+\frac{\Phi_0}{2\pi}\right)\Phi_0+\Phi_{ext}+(L_1+L_3)I^*\right)+L_2\left(-V_g+(R_g+R_{out})I^*\right)}{L_2R_{out}+L_1(R_g+R_{out})}. \tag{SI15}$$

Within our linearized model, the three critical currents $I_c^{(i)}$'s are linear functions of $\Phi_{ext} + m\Phi_0$, and hence correspond to three sets of parallel lines in the $I_{in}$ vs $\Phi_{ext}$ plane. The intersections between these lines determine the vertices of a zigzag pattern, which separates the superconducting region of the device from the normal region. In the superconducting region there exists a value of $m$, for which all internal currents satisfy $I_i < I^*$. On the other hand, in the normal region, at least one of the $I_i$'s is larger than its critical value $I_i^*$.

For $I_1 > I_2$ (or equivalently $I_g < 0$), equations (SI13-SI15) predict a shift of the interference pattern in parallel to the $\Phi_{ext}$ axis. Likewise, for $I_1 < I_2$, these equations predict both a phase and an amplitude shift. In the former case, the superconducting-to-normal transition takes place when the input current reaches the lowest between equations (SI13) and (SI15) ($I_{in} = \min\{I_c^{(1)}, I_c^{(3)}\}$). Remarkably, in both expressions the external magnetic flux $\Phi_{ext}$ and the voltage bias $V_g$ are introduced merely by the expression $\Phi_{ext} - L_2V_g/(R_g + R_{out})$. Thus, the external bias keeps the shape of the interference pattern unchanged, while shifting this pattern along the $\Phi_{ext}$ axis. This phase shift (following Fig. 1d), is given by:

$$\delta\Phi = \frac{L_2V_g}{R_g+R_{out}} = L_2\delta I_g. \tag{SI16}$$

Here, the last identity was derived from Eq. (SI7) by assuming a constant input current, $I_{in}$. If we consider a fixed point on the zigzag pattern (e.g. one of the maxima or minima), we can substitute Eq. (SI16) into Eq. (SI11) to obtain $\delta I_2 = \delta I_g$. This result has a simple physical meaning: at a fixed point along the zigzag pattern, $I_{in}$ is constant and one of the following two conditions is satisfied: $I_1 = I_1^* = I^*$ or $I_3 = I_3^* = -I^*$.

Because $I = I_1 - I_3$ both $I_1$ and $I_3$ are constants and the gate current flows to merge completely into $I_2$. Consequently, the Fulton phase of the device changes by $\delta\Theta = L_2 \delta I_g$.

In devices with a narrow gate ($w_g < w_j$), introducing $I_g$ reduces the critical current of the bottom weak links. We model this critical-current reduction by the modified threshold conditions: $I_1 = I^* - \alpha I_g$ and $I_2 = I^* - \alpha I_g$. These conditions are met when the input current $I_{\text{in}}$ reaches the following two corresponding critical values:

$$I_c^{(1)} = \frac{(\alpha L_1 + (-1+\alpha)L_2 + \alpha L_3)V_g + (R_g + R_{\text{out}})\left(\left(m + \frac{\Theta_0}{2\pi}\right)\Phi_0 + \Phi_{\text{ext}}\right) - (L_1 + L_2 + L_3)(R_g + R_{\text{out}})I^*}{-L_3 R_g + (\alpha L_1 + (-1+\alpha)(L_2 + L_3))R_{\text{out}}}, \quad \text{(SI17)}$$

$$I_c^{(2)} = \frac{((1+\alpha)L_1 + \alpha L_2 + (1+\alpha)L_3)V_g + (R_g + R_{\text{out}})\left(\left(m + \frac{\Theta_0}{2\pi}\right)\Phi_0 + \Phi_{\text{ext}}\right) - (L_1 + L_2 + L_3)(R_g + R_{\text{out}})I^*}{-L_3 R_g + ((1+\alpha)L_1 + \alpha(L_2 + L_3))R_{\text{out}}}. \quad \text{(SI18)}$$

Note that for $\alpha = 0$, equations (SI17) and (SI18) are reduced to equations (SI13) and (SI14), respectively. As mentioned above, amplitude shifts occur for $I_1 > I_2$. That is, when the interference pattern is set by the intersection between the solutions for $I_c^{(2)}$ and $I_c^{(3)}$. Using equations (SI18) and (SI15), we find that the critical current oscillates between

$$\max(I_c) = \frac{(1+\alpha)V_g - 2(R_g + R_{\text{out}})I^*}{-R_g + \alpha R_{\text{out}}} \quad \text{(SI19)}$$

and

$$\min(I_c) = \frac{(R_g + R_{\text{out}})\Phi_0 - (L_1 + L_2 + L_3)\left((1+\alpha)V_g - 2(R_g + R_{\text{out}})I^*\right)}{(L_1 + L_2 + L_3)(R_g - \alpha R_{\text{out}})} \quad \text{(SI20)}$$

Thus, the shifts of the interference pattern along the current axis (i.e. amplitude shift, following Fig. 1d) depends on the voltage as follows:

$$\delta I_c = -\frac{1+\alpha}{\alpha R_{\text{out}} - R_g} V_g, \quad \text{(SI21)}$$

as appears in the main text (Eq. (4)).